\documentclass[preprint2]{aastex6}
\usepackage{epsfig}

\shorttitle{JVLA MCV Survey Data}
\shortauthors{Barrett et al.}
\begin{document}

\title{A Jansky VLA Survey of Magnetic Cataclysmic Variable Stars:
  \\ I. The Data}

\author{Paul E. Barrett\altaffilmark{1,2} and Christopher Dieck}
\affil{United States Naval Observatory \\
  3450 Massachusetts Ave NW \\
  Washington, DC 20392-5420}

\author{Anthony J. Beasley}
\affil{National Radio Astronomical Observatory \\
  520 Edgemont Road \\
  Charlottesville, VA 22903-2475}

\author{Kulinder P. Singh}
\affil{Tata Institute of Fundamental Research \\
  Dr. Homi Bhabha Road, Navy Nagar, Colaba \\
  Mumbai 400005, India}

\author{Paul A. Mason}
\affil{New Mexico State University \\
  PO Box 30001 \\
  Las Cruces, NM 88003-8001}

\altaffiltext{1}{George Washington University}
\altaffiltext{2}{paul.barrett@usno.navy.mil}

\begin{abstract}

The Jansky Very Large Array was used to observe 121 magnetic
cataclysmic variables (MCVs).  We report radio detections of 19 stars.
Fourteen are new radio sources, increasing the number of MCVs that are
radio sources by more than twofold, from 8 to 22.  Most detections are
at 8.7 GHz (X-band) with a lesser number at 5.4 and 21.1 GHz (C- and
K-bands).  Most flux density limits are in the range of 47--470
$\mu$Jy.  With the exception of AE~Aqr, the maximum flux detected is
818 $\mu$Jy.  Fourteen of the detections show approximately 100\%
circularly polarized emission, which is characteristic of
electron-cyclotron maser emission.  The data suggest that MCVs might
be divided into two classes of radio emitters: those dominated by
weakly polarized gyrosynchrotron emission and those by highly
polarized electron-cyclotron maser emission.

\end{abstract}


\section{Introduction}


Cataclysmic variables (CVs) are binary stars consisting of an
accreting white dwarf primary and a lower main sequence, Roche-lobe
filling, secondary.  They are roughly divided into two classes based
on their magnetic properties: systems that have a primary with a
strong ($>1$ MG) magnetic field, the magnetic CVs (MCVs) and those
without, the normal or non-magnetic CVs.  The MCVs are further divided
into two subclasses; the polars (for polarized stars), characterized
by a single optical and X-ray photometric period and strong optical
linear (5--10\%) and circular (10--80\%) polarization, and the
intermediate polars (IPs) and DQ~Her stars, characterized by multiple
optical, and possibly X-ray, photometric periods.  The magnetic field
in the polars is sufficiently strong ($>10$ MG) to completely disrupt
the formation of an accretion disk, whereas the field in the IPs and
DQ~Her stars is inferred to be weaker (1--10 MG) resulting in a
truncated (inner) accretion disk.  The IPs are slow rotators with
P$_{spin}$ = 30--60 minutes and the three DQ~Her systems (DQ~Her,
AE~Aqr, and V533~Her) are fast rotators with P$_{spin}$ = 30--60
seconds, presumably caused by an episode of very high accretion
similar to that of the millisecond pulsars.


Studies of CV evolution in the mid-1980s (\citeauthor{rapp83}
\citeyear{rapp83}, \citeauthor{spru83} \citeyear{spru83}) showed that
the loss of orbital angular momentum due to gravitational radiation at
orbital periods $>2$~h was insufficient to drive mass transfer.
Another mechanism such as magnetic braking was needed.  In this
scenario, the outflowing stellar wind from the magnetized secondary
star generates a braking torque on the binary and a loss of orbital
angular momentum that is sufficiently strong to drive mass transfer at
orbital periods between 3--6~h.  At 3~h the secondary star becomes
fully convective, which effectively turns off the magnetic braking and
mass transfer.  At an orbital period of about 2 h, angular momentum
loss due to gravitational radiation becomes significant and mass
transfer resumes.  This evolutionary scenario explains the lack of CVs
between two and three hours, which is called the period gap.

For MCVs, the strong magnetic field of the WD extends out to and
beyond the secondary.  It is therefore possible that some of the mass
flow from the secondary flows out of the system along the field lines.
At a WD radius of $10^{11}$ cm, the magnetic field of a 10 MG WD is
$\approx 500$ G, which equates to a cyclotron frequency of 1.4 GHz.
Assuming a typical accretion rate of $10^{15}$ g s$^{-1}$, the radio
luminosity is approximately $10^{22}$ ergs $s^{-1}$.  Although this
luminosity was well below the sensitivity of the then current radio
telescopes, \citet{chan78} recommended making radio observations of
these systems.  Based on this recommendation, R.M. Hjellming (personal
communication) observed the polar AM~Her (presumably using the Very
Large Array), but did not detect it at a sensitivity limit of about 1
mJy, where Jy = 10$^{-26}$ W m$^{-2}$ Hz$^{-1}$.


Further observations of AM~Her at 4.9 GHz using the Very Large Array
(VLA) by \citet{chan82} resulted in the first radio detection of a
MCV.  A flux density of 0.67 mJy was observed.  Although the
observations were made in full polarization mode, no circular
polarization was observed.  They concluded that the radio emission
from AM~Her is due to gyrosynchrotron emission (harmonic number
30--50) from energetic ($\approx 350$ keV) electrons trapped in the
magnetosphere of the WD.  EF Eri, which was also observed, was not
detected at a sensitivity limit of $<0.2$ mJy.  They suggest that the
non-detection of EF Eri might be because its magnetic field, which was
not known at the time, is considerably different than AM~Her; or the
orbital separation of the stars may preclude emission at 4.9 GHz.  The
origin of the energetic electrons was not known, but they suggest
several mechanisms for their production, such as shock waves and the
unipolar inductor model of \citet{gold69}.

Subsequent VLA observations of MCVs by \citet{dulk83} again detected
radio emission from AM~Her at 4.9 GHz and gave upper limits at 1.4 and
15 GHz.  They also obtained upper limits of about 0.2 mJy at 4.9 GHz
for VV~Pup, EF~Eri, MR~Ser, ST~LMi, and AN~UMa.  They suggest that the
quiescent emission is the result of energetic electrons ($\approx 500$
keV) trapped in the magnetosphere of the WD, provided that the
electron energy spectrum is quite hard and the spectral hardness or
number density of the energetic electrons increases with radius.
Serendipitously, a radio flare lasting about 10 minutes was observed
during the observation.  The flare had a peak flux of 9.7 mJy
($\approx 20$ times the quiescent emission) and was essentially 100\%
circularly polarized.  They concluded that the radio emission was due
to an electron-cyclotron maser that operates near the surface of the
red-dwarf companion in a $\approx 1$ kG magnetic field.  Magnetic
reconnection events due to the interaction of the magnetic fields of
the two stars may be the source of the energetic electrons that
generate the maser emission.


The first detection of radio emission from a DQ~Her binary was
reported by \citet{book87} who detect strong radio emission ($\approx
15$ mJy at 4.9 GHz and $\approx 5$ mJy at 1.4 GHz) from AE~Aqr using
the VLA.  No radio emission was observed from five other DQ~Her stars
and intermediate polars (FO~Aqr, AO~Psc, BG~CMi, TV~Col, \& EX~Hya).
They obtained upper limits of $\approx 0.2$ mJy at 1.4 and 4.9 GHz.
In an attempt to test the unipolar inductor model, the asynchronous
polar BY~Cam was observed using the VLA \citep{maso96}. The upper
limits obtained for this long duration VLA observation are similar to
the detections of the current work. \citet{maso07}, during a survey of
9 MCVs at 8.4 GHz, also detected AR~UMa at flux density of $\approx
0.6$ mJy.  AR~UMa has the highest known magnetic field of all polars
at 230 MG.  The flux densities of two detections were 422 and 734
$\mu$Jy.  AM~Her was also detected at a flux density of 584 $\mu$Jy.
The other seven targets (LW~Cam, DO~Dra, FIRST J1023.8+0038, SDSS
J1553+5516, V2301~Oph, RX J1846.9+5538, and WZ~Sge) were not not
detected at a flux limit of $\approx 120$ $\mu$Jy.


\citet{bast87} observed 15 MCVs, 8 polars and 7 intermediate polars
and a DQ~Her star, using the VLA.  No detections were made at 1.5,
4.9, and 15 GHz.  \citet{beas94} used the Very Large Array and the
Australian Telescope Compact Array to observe an additional 22 MCVs
without success.  However, \citet{wrig88} reported a detection of the
polar V834 Cen at 8.4 GHz using the Parkes 64m telescope, and
\citet{pave94}, observing 21 polars and IPs using the Jodrell Bank
broadband interferometer, reported detections of five MCVs (BG~CMi,
ST~LMi, DQ~Her, AM~Her, AE~Aqr), three of which were new.  For a
review of the early radio observations of CVs of all types and a
discussion of the radio mechanisms see \citet{chan87}.

Since those early observations, there has been little interest in
observing MCVs in the radio and hence, little progress on
understanding the location and mechanism of the radio emission,
because the few surveys that were done had little success.  However,
during the last five years, the much greater sensitivity of the Jansky
Very Large Array (JVLA) has significantly improved the situation.
\citet{copp15} and \citet{copp16} detected radio emission from dwarf
nova and nova-like CVs at the tens of $\mu$Jy level.  For the first
survey of nova-like CVs, they observed four targets and detected three
of them (RW~Sex, V603~Aql, and TT~Ari) at 6 GHz.  For the second
survey of dwarf novae, they observed five targets (Z~Cam, RX~And,
SU~UMa, YZ~Cnc, and U~Gem) and detected all of them at flux densities
between 12 and 58 $\mu$Jy.  The observed radio emission was variable
on timescales of minutes to days, which is consistent with previous
radio observations of other CVs.

Beginning in 2013, a project was initiated using the Very Long
Baseline Array (VLBA) to perform astrometry of radio bright ($>1$ mJy)
MCVs in an attempt to accurately measure their distances.  These
results will be presented in a separate paper.  At the same time, a
survey was begun using the JVLA to identify any new radio bright MCVs
for follow-up astrometry.  The results of that survey are presented
here and in a subsequent paper.  The current paper presents the
results of the data analysis and briefly discusses some observational
statistics of the survey, while the second paper presents a more
detailed discussion of the data.

\section{Observations and Data Reduction}

We used the JVLA to observe 121 MCVs during two observing semesters,
13B and 15A, at primarily three frequencies (C-, X-, \& K-bands; 4--6,
8--10 \& 18--22 GHz, respectively) at full polarization.  A fourth
frequency, Q-band (40--44 GHz), was also used for a few observations.
Accurate flux densities were obtained by observing one of four
standard flux calibrators (3C48, 3C138, 3C147, and 3C286) during each
scheduling block (SB), except for six SBs during semester 13B that used
3C295 as a flux calibrator and could not be accurately flux calibrated
by the CASA calibration pipeline.  The flux calibrators also served as
linear polarization standards.  However, accurate linear polarization
measurements were not possible, because no polarization leakage
calibrators were observed.  This omission is not expected to
signifanctly impact the measurement of circular polarization.  For
semester 13B, 40~h of observing time was requested to observe 60 of
the optically brightest, and likely closest, MCVs.  The observing
program allowed about three MCVs to be observed in each one hour SB.
Exposures were approximately two minutes per frequency.  Each SB was
scheduled twice to increase the probability of detecting the MCVs.
The JVLA scheduled 35 of the 40~h, resulting in 42 MCVs being
observed.  For semester 15A, 69~h of observing time was requested to
observe 69 MCVs.  Except for one SB, each SB contained two MCVs with
each exposure being approximately five minutes per frequency.  The
JVLA scheduled all 69~h resulting in an additional 69 MCVs being
observed.  These observations are summarized in Table 1.  Column 1
gives the General Catalog of Variable Stars (GCVS) name.  If no GCVS
name is available, the star's abbreviated constellation name followed
by a number is used.  Column 2 is the JVLA semester--scheduling-block
designation.  Column 3 is the date and columns 4--9 are the UT start
times and exposure lengths for the C-, X-, and K-bands, respectively.
\begin{deluxetable*}{lcccccccc}
  \tablewidth{0pt}
  \tabletypesize{\small}
  \tablecaption{Log of JVLA observations}
  \tablehead{
    \colhead{GVCS Name} & \colhead{Semester}         & \colhead{} &
    \multicolumn{2}{c}{C-Band} & \multicolumn{2}{c}{X-Band} & \multicolumn{2}{c}{K-Band} \\
    \colhead{} & \colhead{} & \colhead{Date} &
    \colhead{Start} & \colhead{Expo}   & \colhead{Start} & \colhead{Expo}     & \colhead{Start} & \colhead{Expo} \\
    \colhead{}     & \colhead{}         & \colhead{UT} &
    \colhead{UT}     & \colhead{(s)}     & \colhead{UT} & \colhead{(s)} & \colhead{UT} & \colhead{(s)}
  }
  \startdata
  FL Cet        &  13B-02A  &  2013-09-11  &  11:38:50  &      110  &  11:31:21  &      111  &  11:35:03  &       111 \\
  V1294 Tau     &  13B-02A  &  2013-09-11  &  11:52:25  &      115  &  11:44:09  &      120  &  11:48:18  &       120 \\
  V1309 Ori     &  13B-02A  &  2013-09-11  &  12:04:20  &      110  &  11:56:51  &      108  &  12:00:36  &       111 \\
  V709 Cas      &  13B-01A  &  2013-09-12  &  11:13:05  &      105  &  11:09:36  &      105  &  11:06:06  &       102 \\
  Cas 1         &  13B-01A  &  2013-09-12  &  11:02:05  &  \phn 90  &  10:58:48  &  \phn 96  &  10:55:27  &   \phn 93 \\
  BY Cam        &  13B-01A  &  2013-09-12  &  11:25:15  &      105  &  11:21:42  &      105  &  11:18:15  &       105 \\
  V2301 Oph     &  13B-16A  &  2013-09-13  &  03:52:00  &      135  &  03:44:21  &      135  &  03:36:42  &       135 \\
  V884 Her      &  13B-16A  &  2013-09-13  &  03:31:55  &      130  &  03:27:33  &      132  &  03:23:12  &       135 \\
\enddata
\end{deluxetable*}

All data are calibrated using CASA version 4.2.2 and the JVLA
calibration pipeline version 1.3.1.  Although scheduling blocks 15--17
are not properly flux calibrated because CASA does not have models for
the flux standard, 3C295; the flux density estimates of these data do
not appear to be grossly different compared to the properly calibrated
data.  Note that an incorrect flux calibration does not affect the
presence of a radio source.  The resulting calibrated images were then
cleaned using an automated script.  An image size of $2500 \times
2500$ pixels was used at a scale of 0.02 arcseconds per pixel,
resulting in a $50 \times 50$ arcsecond image.  Six sets of uncleaned
images were created; specifically, I, Q, U, V, LL, and RR.  Natural
weighting was used for all images.  For a two minute integration, RMS
errors of 25, 25, and 33 $\mu$Jy/beam are expected for the C-, X-, and
K-bands, respectively.

\section{Results}


Table 2 lists the 121 MCVs observed during this survey and their CV
subclass and optical positions.  As in Table 1, column 1 is the GCVS
name.  Column 2 is the CV subclass.  Columns 3 and 4 are the right
ascension (RA) and declination (Dec) of the star at epoch J2000.
Columns 5 and 6 are the proper motion of the star, if known.  Proper
motions are primarily from the US Naval Observatory Combined
Astrometric Catalog (UCAC4) supplemented by the Simbad on-line catalog
when necessary.  Columns 7 and 8 are the RA and Dec of the star at the
epoch of the observation.  The last two columns enable a comparison of
the observed position to the cataloged position.
\begin{deluxetable*}{llrrrrrr}
  \tablewidth{0pt}
  \tabletypesize{\small}
  \tablecaption{Target list and their properties}
  \tablehead{
    \colhead{GCVS Name}  & \colhead{Class\tablenotemark{a}} & \colhead{RA} & \colhead{Dec} &
    \colhead{PM RA} & \colhead{PM Dec} & \colhead{RA+PM} & \colhead{Dec+PM} \\
    \colhead{}  & \colhead{}      & \colhead{(hh mm ss.ss)} & \colhead{(dd mm ss.s)} &
    \colhead{(mas)} & \colhead{(mas)}  & \colhead{(hh mm ss.ss)} & \colhead{(dd mm ss.s)}
  }
  \startdata
  V709 Cas     & IP   & 00 28 48.84 & +59 17 22.3 &    0.2 &   -1.8 & 00 28 48.84 & +59 17 22.3 \\
  And 1        & IP ? & 00 43 08.61 & +41 12 47.6 &    7.1 &    1.5 & 00 43 08.62 & +41 12 47.6 \\
  EQ Cet       & AM   & 01 28 52.52 & -23 39 43.8 &     29 &    -34 & 01 28 52.55 & -23 39 44.3 \\
  Cas 1        & IP   & 01 53 20.76 & +74 46 22.2 &        &        & 01 53 20.76 & +74 46 22.2 \\
  FL Cet       & AM   & 01 55 43.44 & +00 28 06.4 &   64.3 &   13.9 & 01 55 43.50 & +00 28 06.6 \\
  AI Tri       & AM   & 02 03 48.60 & +29 59 26.0 &   -8.5 &  -16.2 & 02 03 48.59 & +29 59 25.8 \\
  TT Ari       & VY   & 02 06 53.09 & +15 17 41.8 &     -8 &  -24.2 & 02 06 53.08 & +15 17 41.5 \\
  BS Tri       & AM   & 02 09 29.80 & +28 32 29.3 &        &        & 02 09 29.80 & +28 32 29.3 \\
  VW For       & AM ? & 02 52 51.35 & -30 37 42.6 &    8.6 &  -30.1 & 02 52 51.36 & -30 37 43.1 \\
  IQ Eri       & DN   & 02 55 38.03 & -22 47 02.7 &     29 &   -1.9 & 02 55 38.06 & -22 47 02.7 \\
  XY Ari       & IP   & 02 56 08.19 & +19 26 34.1 &        &        & 02 56 08.19 & +19 26 34.1 \\
  EF Eri       & AM   & 03 14 13.26 & -22 35 43.3 &    123 &    -47 & 03 14 13.38 & -22 35 44.0 \\
  GK Per       & IP   & 03 31 12.01 & +43 54 15.4 &     -8 &  -17.9 & 03 31 12.00 & +43 54 15.2 \\
  VY For       & AM   & 03 32 04.61 & -25 56 55.0 &        &        & 03 32 04.61 & -25 56 55.0 \\
  UZ For       & AM   & 03 35 28.62 & -25 44 22.3 &        &        & 03 35 28.62 & -25 44 22.3 \\
  V1294 Tau    & VY   & 04 00 37.23 & +06 22 46.3 &    2.9 &   -4.4 & 04 00 37.23 & +06 22 46.2 \\
  AH Eri       & IP ? & 04 22 38.05 & -13 21 30.5 &        &        & 04 22 38.05 & -13 21 30.5 \\
  IW Eri       & AM   & 04 25 55.18 & -19 45 30.4 &        &        & 04 25 55.18 & -19 45 30.4 \\
  HY Eri       & AM   & 05 01 46.39 & -03 59 20.7 &        &        & 05 01 46.39 & -03 59 20.7 \\
  V1062 Tau    & IP   & 05 02 27.47 & +24 45 23.4 &   -4.1 &    0.8 & 05 02 27.47 & +24 45 23.4 \\
  Tau 4        & AM ? & 05 02 50.94 & +16 24 22.0 &    2.9 &     -5 & 05 02 50.94 & +16 24 21.9 \\
  UU Col       & IP   & 05 12 13.22 & -32 41 39.8 &   -2.2 &     11 & 05 12 13.22 & -32 41 39.6 \\
  V1309 Ori    & AM   & 05 15 41.41 & +01 04 40.4 &    4.1 &   -8.2 & 05 15 41.41 & +01 04 40.3 \\
  Aur (Paloma) & AM   & 05 24 30.48 & +42 44 50.2 &    4.2 &      7 & 05 24 30.48 & +42 44 50.3 \\
  TV Col       & IP   & 05 29 25.52 & -32 49 04.0 &   13.3 &   16.9 & 05 29 25.53 & -32 49 03.8 \\
  BY Cam       & AM   & 05 42 48.77 & +60 51 31.5 &    -41 &   10.5 & 05 42 48.73 & +60 51 31.6 \\
  LS Cam       & IP ? & 05 57 23.93 & +72 41 52.6 &   -3.8 &    9.4 & 05 57 23.93 & +72 41 52.7 \\
  V405 Aur     & IP   & 05 57 59.27 & +53 53 45.1 &   -2.6 &  -12.9 & 05 57 59.27 & +53 53 44.9 \\
  Lep 1        & AM ? & 06 00 33.30 & -27 09 18.5 &     -5 &    0.1 & 06 00 33.29 & -27 09 18.5 \\
  V552 Aur     & IP ? & 06 14 09.81 & +45 30 08.5 &    7.7 &   -7.3 & 06 14 09.82 & +45 30 08.4 \\
  MU Cam       & IP   & 06 25 16.18 & +73 34 39.2 &    3.5 &    4.5 & 06 25 16.18 & +73 34 39.3 \\
  V647 Aur     & IP   & 06 36 32.55 & +35 35 43.3 &    1.5 &    0.9 & 06 36 32.55 & +35 35 43.3 \\
  Mon 1        & AM   & 06 49 50.88 & -07 37 41.7 &    6.7 &    8.7 & 06 49 50.89 & -07 37 41.6 \\
  V418 Gem     & IP   & 07 04 08.67 & +26 25 10.9 &    1.9 &    1.8 & 07 04 08.67 & +26 25 10.9 \\
  LW Cam       & AM   & 07 04 09.98 & +62 03 28.3 &    9.7 &  -12.9 & 07 04 09.99 & +62 03 28.1 \\
  V348 Pup     & SW   & 07 12 32.90 & -36 05 38.6 &    1.4 &    8.1 & 07 12 32.90 & -36 05 38.5 \\
  HS Cam       & AM   & 07 19 14.49 & +65 57 45.0 &        &        & 07 19 14.49 & +65 57 45.0 \\
  GI Mon       & IP ? & 07 26 47.10 & -06 40 29.5 &      7 &    5.4 & 07 26 47.11 & -06 40 29.4 \\
  BG CMi       & AM   & 07 31 29.00 & +09 56 23.1 &    -15 &  -21.1 & 07 31 28.99 & +09 56 22.8 \\
  Mon 2        & AM   & 07 49 10.40 & -05 49 25.6 &   -1.5 &    3.3 & 07 49 10.40 & -05 49 25.5 \\
  PQ Gem       & IP   & 07 51 17.33 & +14 44 23.9 &   -4.3 &    7.7 & 07 51 17.33 & +14 44 24.0 \\
  EU Lyn       & AM   & 07 52 40.45 & +36 28 23.2 &   -0.4 &    5.5 & 07 52 40.45 & +36 28 23.3 \\
  EV Lyn       & SW ? & 07 54 43.01 & +50 07 29.2 &     -3 &   -5.9 & 07 54 43.01 & +50 07 29.1 \\
  HT Cam       & IP   & 07 57 01.37 & +63 06 01.8 &   -3.4 &   -1.1 & 07 57 01.37 & +63 06 01.8 \\
  DW Cnc       & IP   & 07 58 53.03 & +16 16 45.2 &  -20.1 &   -3.4 & 07 58 53.01 & +16 16 45.1 \\
  Cnc 1        & AM ? & 07 59 39.79 & +19 14 17.3 &        &        & 07 59 39.79 & +19 14 17.3 \\
  V351 Pup     & IP ? & 08 11 38.40 & -35 07 30.5 &        &        & 08 11 38.40 & -35 07 30.5 \\
  VV Pup       & AM   & 08 15 06.78 & -19 03 17.8 &    9.7 &  -70.7 & 08 15 06.79 & -19 03 18.8 \\
  EG Lyn       & AM   & 08 20 51.07 & +49 34 31.7 &        &        & 08 20 51.07 & +49 34 31.7 \\
  WX Pyx       & IP   & 08 33 05.74 & -22 48 32.2 &   -0.6 &    1.2 & 08 33 05.74 & -22 48 32.2 \\
  SW UMa       & SU   & 08 36 42.74 & +53 28 38.1 &        &        & 08 36 42.74 & +53 28 38.1 \\
  Lyn 1        & AM   & 08 37 51.00 & +38 30 12.5 &   -0.6 &    7.2 & 08 37 51.00 & +38 30 12.6 \\
  Cnc 2        & IP ? & 08 46 17.12 & +24 34 44.1 &        &        & 08 46 17.12 & +24 34 44.1 \\
  EU Cnc       & AM   & 08 51 27.19 & +11 46 57.0 &  -13.8 &   -6.9 & 08 51 27.18 & +11 46 56.9 \\
  Hya 1        & AM   & 08 59 09.18 & +05 36 54.5 &        &        & 08 59 09.18 & +05 36 54.5 \\
  GZ Cnc       & SU   & 09 15 51.67 & +09 00 49.6 &  -36.4 &  -29.8 & 09 15 51.63 & +09 00 49.1 \\
  HU Leo       & AM ? & 09 24 44.48 & +08 01 51.0 &    6.8 &   -9.7 & 09 24 44.49 & +08 01 50.8 \\
  HS 0922+1333 & AM   & 09 24 55.94 & +13 20 52.4 &    -17 &    2.6 & 09 24 55.92 & +13 20 52.4 \\
  MN Hya       & AM   & 09 29 07.04 & -24 05 05.4 &    8.3 &   -0.9 & 09 29 07.05 & -24 05 05.4 \\
  US 691       & SU   & 09 32 49.57 & +47 25 23.0 &        &        & 09 32 49.57 & +47 25 23.0 \\
  VZ Sex       & IP ? & 09 44 31.70 & +03 58 05.6 &  -18.3 &  -17.9 & 09 44 31.68 & +03 58 05.3 \\
  HY Leo       & IP   & 09 46 34.50 & +13 50 58.1 &     -9 &   -1.5 & 09 46 34.49 & +13 50 58.1 \\
  Leo 1        & AM   & 09 53 08.17 & +14 58 36.1 &        &        & 09 53 08.17 & +14 58 36.1 \\
  Hya 2        & AM   & 10 02 11.71 & -19 25 37.4 &        &        & 10 02 11.71 & -19 25 37.4 \\
  Hya 3        & AM   & 10 07 34.64 & -20 17 32.4 &        &        & 10 07 34.64 & -20 17 32.4 \\
  GG Leo       & AM   & 10 15 34.67 & +09 04 42.0 &  -29.3 &   -8.4 & 10 15 34.64 & +09 04 41.9 \\
  WX LMi       & AM   & 10 26 27.49 & +38 45 02.5 &     28 &    -30 & 10 26 27.52 & +38 45 02.0 \\
  YY Sex       & IP ? & 10 39 46.97 & -05 06 58.3 &  -25.7 &     -4 & 10 39 46.96 & -05 00 58.1 \\
  FH UMa       & AM   & 10 47 09.93 & +63 35 13.1 &        &        & 10 47 09.93 & +63 35 13.1 \\
  EK UMa       & AM   & 10 51 35.15 & +54 04 36.1 &        &        & 10 51 35.15 & +54 04 36.1 \\
  AN UMa       & AM   & 11 04 25.67 & +45 03 14.0 &  -28.6 &  -17.9 & 11 04 25.64 & +45 03 13.8 \\
  ST LMi       & AM   & 11 05 39.76 & +25 06 28.7 &   -1.2 &  -35.9 & 11 05 39.76 & +25 06 28.2 \\
  AR UMa       & AM   & 11 15 44.56 & +42 58 22.4 &    -77 &      2 & 11 15 44.49 & +42 58 22.4 \\
  DP Leo       & AM   & 11 17 15.93 & +17 57 42.0 &        &        & 11 17 15.93 & +17 57 42.0 \\
  DO Dra       & IP   & 11 43 38.50 & +71 41 20.7 &   16.9 &   10.9 & 11 43 38.52 & +71 41 20.9 \\
  EU UMa       & AM   & 11 49 55.71 & +28 45 07.6 &        &        & 11 49 55.71 & +28 45 07.6 \\
  EV UMa       & AM   & 13 07 53.87 & +53 51 30.6 &        &        & 13 07 53.87 & +53 51 30.6 \\
  Vir 1        & AM   & 14 22 56.32 & -02 21 08.0 &        &        & 14 22 56.32 & -02 21 08.0 \\
  Vir 2        & AM   & 14 24 38.93 & -02 27 39.3 &        &        & 14 24 38.93 & -02 27 39.3 \\
  BM CrB       & AM   & 15 41 04.67 & +36 02 52.9 &        &        & 15 41 04.67 & +36 02 52.9 \\
  MR Ser       & AM   & 15 52 47.18 & +18 56 29.2 &     11 &   51.7 & 15 52 47.19 & +18 56 29.9 \\
  AP CrB       & AM   & 15 54 12.34 & +27 21 52.4 &    -18 &     24 & 15 54 12.32 & +27 21 52.8 \\
  V519 Ser     & AM   & 16 10 07.51 & +03 52 33.0 &        &        & 16 10 07.51 & +03 52 33.0 \\
  Her 1        & AM   & 16 26 08.16 & +33 28 27.8 &        &        & 16 26 08.16 & +33 28 27.8 \\
  V1189 Her    & AM   & 16 29 36.53 & +26 35 19.6 &    -12 &     -6 & 16 29 36.52 & +26 35 19.5 \\
  V1237 Her    & AM   & 17 00 53.30 & +40 03 57.7 &        &        & 17 00 53.30 & +40 03 57.7 \\
  V795 Her     & IP   & 17 12 56.17 & +33 31 19.2 &      3 &  -23.7 & 17 12 56.17 & +33 31 18.9 \\
  V1007 Her    & AM   & 17 24 06.28 & +41 14 08.0 &   -2.4 &    3.2 & 17 24 06.28 & +41 14 08.0 \\
  V2731 Oph    & IP   & 17 30 21.90 & -05 59 32.1 &      1 &   12.9 & 17 30 21.90 & -05 59 31.9 \\
  Sgr 1        & AM   & 17 45 32.70 & -29 05 52.0 &    0.3 &   -0.6 & 17 45 32.70 & -29 05 52.0 \\
  V2301 Oph    & AM   & 18 00 35.58 & +08 10 13.6 &        &        & 18 00 35.58 & +08 10 13.6 \\
  V884 Her     & AM   & 18 02 06.52 & +18 04 44.9 &  -46.8 &  -75.5 & 18 02 06.48 & +18 04 43.9 \\
  V1323 Her    & IP   & 18 03 39.67 & +40 12 20.6 &    1.3 &   -1.8 & 18 03 39.67 & +40 12 20.6 \\
  DQ Her       & IP   & 18 07 30.26 & +45 51 32.1 &    5.3 &   16.7 & 18 07 30.26 & +45 51 32.3 \\
  V426 Oph     & IP ? & 18 07 51.69 & +05 51 47.8 &    1.7 &  -34.4 & 18 07 51.69 & +05 51 47.3 \\
  AM Her       & AM   & 18 16 13.33 & +49 52 04.3 &  -46.3 &     27 & 18 16 13.29 & +49 52 04.7 \\
  Sct 1        & IP ? & 18 32 20.00 & -08 40 30.0 &  -12.1 &    3.8 & 18 32 19.99 & -08 40 29.9 \\
  MT Dra       & AM   & 18 46 58.83 & +55 38 29.1 &   -9.9 &   -8.8 & 18 46 58.82 & +55 38 29.0 \\
  V603 Aql     & SH ? & 18 48 54.64 & +00 35 02.9 &   10.8 &   -8.7 & 18 48 54.65 & +00 35 02.8 \\
  V373 Sct     & IP ? & 18 55 26.71 & -07 43 05.5 &        &        & 18 55 26.71 & -07 43 05.5 \\
  V1425 Aql    & IP ? & 19 05 26.68 & -01 42 14.2 &    3.5 &    4.4 & 19 05 26.68 & -01 42 14.1 \\
  EP Dra       & AM   & 19 07 06.18 & +69 08 44.2 &   -1.2 &  -10.3 & 19 07 06.18 & +69 08 44.0 \\
  V407 Vul     & AM ? & 19 14 26.07 & +24 56 43.3 &   -9.9 &    0.7 & 19 14 26.06 & +24 56 43.3 \\
  V1432 Aql    & AM   & 19 40 11.42 & -10 25 25.8 &  -15.2 &   -3.5 & 19 40 11.41 & -10 25 25.8 \\
  V2306 Cyg    & IP   & 19 58 14.48 & +32 32 42.2 &    0.9 &      2 & 19 58 14.48 & +32 32 42.2 \\
  QQ Vul       & AM   & 20 05 41.91 & +22 39 58.9 &    9.2 &   -8.2 & 20 05 41.92 & +22 39 58.8 \\
  WZ Sge       & WZ ? & 20 07 36.41 & +17 42 15.4 &    102 &    -10 & 20 07 36.50 & +17 42 15.3 \\
  V4738 Sgr    & AM   & 20 22 37.53 & -39 54 12.8 &        &        & 20 22 37.53 & -39 54 12.8 \\
  AE Aqr       & IP   & 20 40 09.16 & -00 52 15.1 &   68.3 &   14.3 & 20 40 09.23 & +00 52 14.9 \\
  Aqr 1        & AM   & 20 48 27.91 & +00 50 08.9 &    5.9 &    3.4 & 20 48 27.92 & +00 50 09.0 \\
  HU Aqr       & AM   & 21 07 58.22 & -05 17 40.1 &  -61.9 &  -55.1 & 21 07 58.16 & -05 17 40.9 \\
  V1500 Cyg    & AM   & 21 11 36.51 & +48 09 02.8 &    8.1 &   -2.6 & 21 11 36.52 & +48 09 02.8 \\
  V2069 Cyg    & IP   & 21 23 44.84 & +42 18 01.8 &   -1.7 &   -9.6 & 21 23 44.84 & +42 18 01.7 \\
  Cyg 1        & IP   & 21 33 43.65 & +51 07 24.5 &     -2 &    4.5 & 21 33 43.65 & +51 07 24.6 \\
  LS Peg       & IP   & 21 51 57.93 & +14 06 53.3 &   17.9 &  -15.7 & 21 51 57.95 & +14 06 53.1 \\
  V388 Peg     & AM   & 21 57 32.30 & +08 55 15.4 &   -1.5 &   -1.6 & 21 57 32.30 & +08 55 15.4 \\
  FO Aqr       & IP   & 22 17 55.39 & -08 21 03.8 &    4.1 &    1.6 & 22 17 55.39 & -08 21 03.8 \\
  Aqr 2        & IP ? & 22 38 43.83 & +01 08 20.7 &  -20.6 &   -0.3 & 22 38 43.81 & +01 08 20.7 \\
  AO Psc       & IP   & 22 55 17.99 & -03 10 40.0 &    9.3 &  -20.5 & 22 55 18.00 & -03 10 40.3 \\
  Aqr 5        & AM   & 23 16 03.65 & -05 27 08.7 &   -4.4 &  -18.1 & 23 16 03.65 & -05 27 09.0 \\
  BW Scl       & WZ   & 23 53 00.82 & -38 51 46.0 &   74.7 &  -58.7 & 23 53 00.90 & -38 51 46.9
  \enddata
  \tablenotetext{a}{The CV subclasses are: AM = AM Herculis star (or
    polar), a subclass of MCVs; DN = dwarf nova; IP = intermediate
    polar and DQ Herculis star, subclasses of MCVs; SH = non-SU UMa
    star showing permanent or transient superhumps, a subclass of
    nova-like (NL) CVs; SU = SU Ursa Majoris star, a subtype of DN; SW
    = SW Sextans star, a subtype of NLs; VY = VY Sculptoris star, a
    subclass of NLs; WZ = WZ Sagittae star, a subclass of SU UMa
    stars; ? = uncertain classification.}
\end{deluxetable*}

Table 3 is the list of 19 detected MCVs.  As in Table 1, columns 1 and
2 are the GCVS name and semester -- scheduling-block designation.
These two values uniquely determine the observation.  Columns 3 and 4
are the observed source position.  For stars with detections at
multiple frequencies, the most accurate position is given, which
equates to the measured position of the highest frequency observation.
Column 5 is the distance between the observed and cataloged source
positions and its standard deviation is the approximate size of the
longest axis of the synthesized beam at half-width-half-maximum.
Columns 6--8 are respectively the detected flux and error at the C-,
X-, and K-band frequencies.  Column 9 is the significance of the
detection, represented as a signal-to-noise (S/N) value.  The
significance of each observation is a combination of the source
position and flux densities.  Six of the new detections (Cas~1,
BS~Tri, UZ~For, WX~LMi, BM~CrB, and Her~1) have no measured proper
motion, so their cataloged position is for epoch J2000 and not for the
epoch of the observation (i.e., circa J2014).  This likely explains
the discrepancy between the observed and catalog positions for some of
these stars.  As for V1007~Her, AM~Her, and QQ~Vul that have proper
motions, we currently have no explanation for the large discrepancy in
the observed and cataloged positions.  
\begin{deluxetable*}{llcccrrrrc}
  \tablewidth{0pt}
  \tabletypesize{\small}
  \floattable
  \rotate
  \tablecaption{Radio fluxes and positions of detected MCVs}
  \tablehead{
    \colhead{GVCS Name}   & \colhead{Semester} & \multicolumn{2}{c}{Observed Location}       & \colhead{Distance\tablenotemark{a}} &
    \colhead{C Flux}    & \colhead{X Flux}    & \colhead{K Flux}    & \colhead{S/N}  & \colhead{Notes\tablenotemark{b}\tablenotemark{c}}  \\
    \colhead{}   & \colhead{}        & \colhead{RA}     & \colhead{Dec}            & \colhead{$\Delta d$}                &
    \colhead{Density}   & \colhead{Density}   & \colhead{Density}   & \colhead{}     & \colhead{}  \\
    \colhead{}       & \colhead{}        & \colhead{(hh mm ss)} & \colhead{(dd mm ss)} & \colhead{(arcsec)}                        &
    \colhead{($\mu$Jy)} & \colhead{($\mu$Jy)} & \colhead{($\mu$Jy)} & \colhead{}     & \colhead{}   \\
  }
  \startdata
  EQ Cet       &  15A-01A  &  01 28 52.633  &  -23 39 44.93  & $ 0.6 \pm 0.2$ &              &              & $404 \pm 67$ &   6.1  & RCP  \\
  Cas 1        &  13B-01A  &  01 53 21.091  &  +74 46 23.96  & $ 5.3 \pm 1.6$ & $228 \pm 42$ &              &              &   5.5  & LCP, no-PM  \\
  BS Tri       &  15A-02A  &  02 09 29.820  &  +28 32 28.79  & $ 0.6 \pm 0.6$ &              & $ 66 \pm 21$ &              &   3.4  & no-PM \\
  EF Eri       &  13B-04A  &  03 14 13.398  &  -22 35 43.25  & $ 0.8 \pm 1.0$ &              & $352 \pm 37$ &              &   9.6  & LCP  \\
  UZ For       &  15A-03A  &  03 35 28.727  &  -25 44 22.66  & $ 0.3 \pm 1.5$ & $252 \pm 24$ &              &              &  11.7  & LCP, no-PM  \\
  Tau 4        &  15A-05A  &  05 02 50.981  &  +16 24 21.42  & $ 0.8 \pm 0.5$ &              & $250 \pm 50$ &              &   5.1  & LCP  \\
  VV Pup       &  13B-08B  &  08 15 06.843  &  -19 03 19.26  & $ 0.1 \pm 0.3$ &              & $161 \pm 23$ & $203 \pm 29$ &  10.5  & RCP  \\
  WX LMi       &  15A-18A  &  10 26 27.517  &  +38 45 02.00  & $ 0.7 \pm 0.2$ & $ 93 \pm 15$ & $115 \pm 15$ &              &  10.0  & no-PM  \\
  ST LMi       &  13B-10B  &  11 05 39.765  &  +25 06 28.24  & $ 0.0 \pm 0.4$ &              & $371 \pm 46$ &              & $>50.0$ & LCP  \\
  AR UMa       &  13B-11A  &  11 15 44.488  &  +42 58 22.41  & $ 0.0 \pm 0.2$ & $378 \pm 16$ & $420 \pm 18$ & $467 \pm 35$ & $>50.0$ &  \\
  AR UMa       &  13B-11B  &  11 15 44.487  &  +42 58 22.41  & $ 0.0 \pm 0.2$ & $539 \pm 22$ & $461 \pm 18$ & $352 \pm 33$ & $>50.0$ &  \\
  BM CrB       &  13B-13A  &  15 41 04.666  &  +36 02 52.69  & $ 0.2 \pm 0.4$ &              & $158 \pm 28$ &              &   6.1  & LCP, no-PM  \\
  MR Ser       &  13B-13A  &  15 52 47.153  &  +18 56 29.68  & $ 0.5 \pm 0.4$ & $603 \pm 35$ & $424 \pm 27$ &              &  23.6  & RCP  \\
  Her 1        &  15A-24A  &  16 26 08.212  &  +33 28 27.00  & $ 1.1 \pm 0.2$ &              &              & $218 \pm 22$ &   9.9  & RCP, no-PM  \\
  V1007 Her    &  15A-25A  &  17 24 06.299  &  +41 14 09.97  & $ 2.0 \pm 0.8$ &              & $165 \pm 25$ &              &   6.6  & RCP  \\
  V1323 Her    &  15A-25A  &  18 03 39.666  &  +40 12 19.61  & $ 1.0 \pm 0.6$ &              & $147 \pm 23$ &              &   6.5  & LCP  \\
  V1323 Her    &  15A-25B  &  18 03 39.545  &  +40 12 18.71  & $ 2.8 \pm 1.4$ & $121 \pm 20$ &              &              &   6.1  & RCP  \\
  AM Her       &  13B-15A  &  18 16 13.187  &  +49 52 05.29  & $ 1.8 \pm 1.0$ & $114 \pm 12$ &              &              &   9.5  &  \\
  AM Her       &  13B-15B  &  18 16 13.186  &  +49 52 05.15  & $ 1.4 \pm 0.4$ & $154 \pm 10$ & $243 \pm 10$ & $818 \pm 92$ & $>50.0$ &  \\
  V603 Aql     &  13B-17A  &  18 48 54.647  &  +00 35 02.88  & $ 0.3 \pm 1.5$ &              & $ 59 \pm 19$ &              &   6.0  & LCP  \\
  V603 Aql     &  13B-17B  &  18 48 54.668  &  +00 35 03.01  & $ 0.1 \pm 1.4$ & $ 47 \pm  8$ & $ 88 \pm  8$ &              &  18.9  & RCP  \\
  QQ Vul       &  13B-18B  &  20 05 41.901  &  +22 40 01.89  & $ 3.1 \pm 0.2$ &              &              & $438 \pm 43$ &  10.2  & LCP \\
  AE Aqr       &  15A-30A  &  20 40 09.231  &  -00 52 14.86  & $ 0.0 \pm 0.2$ &$5000 \pm 22$ &$5472 \pm 18$ &$8354 \pm 38$ & $>50.0$ &  \\
  AE Aqr       &  15A-30B  &  20 40 09.233  &  -00 52 14.86  & $ 0.0 \pm 0.2$ &$5041 \pm 20$ &$5211 \pm 16$ &$4771 \pm 30$ & $>50.0$ &
  \enddata
  \tablenotetext{a}{$\Delta d$ is the distance between the observed
    and cataloged source positions and its standard deviation is the
    approximate \\
    size of the longest axis of the synthesize beam at
    half-width-half-maximum.}
  \tablenotetext{b} {LCP and RCP denotes that the detection is seen in
    only the left-hand or right-hand circular polarization channel.}
  \tablenotetext{c} {no-PM denotes that the source has no known proper
    motion.  Therefore, the distance between the observed and \\
    cataloged positions may be large.}  
\end{deluxetable*}

To estimate the probability that an unrelated extragalactic background
radio source close to the stellar position may lead to a false
detection, we use the radio source counts statistics of
\citet{cond12}.  At an observing frequency of 3 GHz the number of
extragalactic backgrounds sources above 100 $\mu$Jy is approximately
1.15 million per steradian.  This corresponds to an areal density of
$2.7 \times 10^{-5}$ per square arcsecond, i.e. the chances of a
unrelated background source above 100 $\mu$Jy being within a radius of
an arcsecond of our stellar positions is extremely low, approximately
1 in 12,000.  Assuming a standard spectral index for the background
sources falling as spectral index $\nu^{-0.6}$, and ignoring unrelated
phenomena such as interferometry resolution effects and source
variability, we can also derive the chances of a confusing source
within 1 arcsecond radius from the stellar positions above 100 $\mu$Jy
at 8 and 22 GHz (1 in 21,000 and 1 in 37,000, respectively).
 
These statistics indicate that extragalactic background sources are
unlikely to contaminate our stellar radio detections.  Systematic
effects such as grating sidelobes from nearby sources distributing
apparent radio flux across our detection images, and the effects of
extended thermal and non-thermal emission in some directions close to
the Galactic plane, can lead to apparent radio emission from our
stellar positions for some JVLA configurations.  In these situations
additional tests (seeking point sources, including CLEANing larger
fields to remove sidelobes, comparing tentative stellar emission
detections amongst different intermediate frequency and polarization
channels, and developing an understanding of the regions our stellar
sources are located in) can be used to develop confidence in a
detection.

Table 4 lists the maximum flux density and its standard deviation for
all observations, i.e., for detections and non-detections.  A region
having a diameter of 4 arcseconds and centered on the expected source
location is used unless the observation contains a detection.  In that
case a background region not affected by any radio sources was used.
Columns 1 and 2 are the same as those in Table 3.  Columns 3--8 are
the maximum flux density and its standard deviation for the C-, X-,
and K-bands, respectively.  For nondetections, Table 4 provides an
estimate of the flux density upper limit.
\begin{deluxetable*}{lcrrr}
  \tablewidth{0pt}
  \tabletypesize{\small}
  \tablecaption{Image Statistics}
  \tablehead{
    \colhead{GCVS Name} & \colhead{Semester} & \colhead{C Flux}    & \colhead{X Flux}    & \colhead{K Flux}    \\
    \colhead{} & \colhead{}        & \colhead{Density}   & \colhead{Density}   & \colhead{Density}   \\
    \colhead{}     & \colhead{}        & \colhead{($\mu$Jy/beam)} & \colhead{($\mu$Jy/beam)} & \colhead{($\mu$Jy/beam)} \\
  }
  \startdata
  V709 Cas      &  13B-01A  & $  75 \pm   35$  & $  71 \pm   26$  & $ 212 \pm   86$ \\
  V709 Cas      &  13B-01B  & $  56 \pm   26$  & $  52 \pm   18$  & $  99 \pm   34$ \\
  And 1         &  15A-32A  & $  26 \pm   11$  & $  46 \pm   12$  & $ 171 \pm   35$ \\
  And 1         &  15A-32B  & $  48 \pm   14$  & $  49 \pm   14$  & $ 123 \pm   33$ \\
  EQ Cet        &  15A-01A  & $  40 \pm   18$  & $  30 \pm   12$  & $ 173 \pm   36$ \\
  EQ Cet        &  15A-01B  & $  48 \pm   14$  & $  47 \pm   12$  & $ 194 \pm   38$ \\
  Cas 1         &  13B-01A  & $ 104 \pm   27$  & $  51 \pm   19$  & $ 263 \pm   84$ \\
  Cas 1         &  13B-01B  & $  50 \pm   22$  & $  54 \pm   19$  & $ 128 \pm   38$ \\
\enddata
\end{deluxetable*}

\section{Discussion}





There are currently eleven known CVs that are radio sources: four are
polars, AM~Her, AR~UMa, ST~LMi, and V834~Cen; four are IPs or DQ~Her
stars, AE~Aqr, BG~CMi, DQ~Her, and GK~Per; and three are considered
non-magnetic CVs: RW~Sex, TT~Ari, and V603~Aql, which are classified
as dwarf novae or nova-likes.  This survey increases the number of CVs
that are radio sources to 25.  The significance of all detections
except one (BS Tri) are ($>5 \sigma$).  This result is a twofold
increase in the number of CVs that are radio sources.  Fifteen of the
new detections are polars; three are IPs, and one is a non-magnetic
CV.



The radio fluxes of most detections are in the range 47--470 $\mu$Jy,
except for AE~Aqr, AM~Her, AR~UMa, and MR Ser (see Figure 1).
\begin{figure}
  \fig{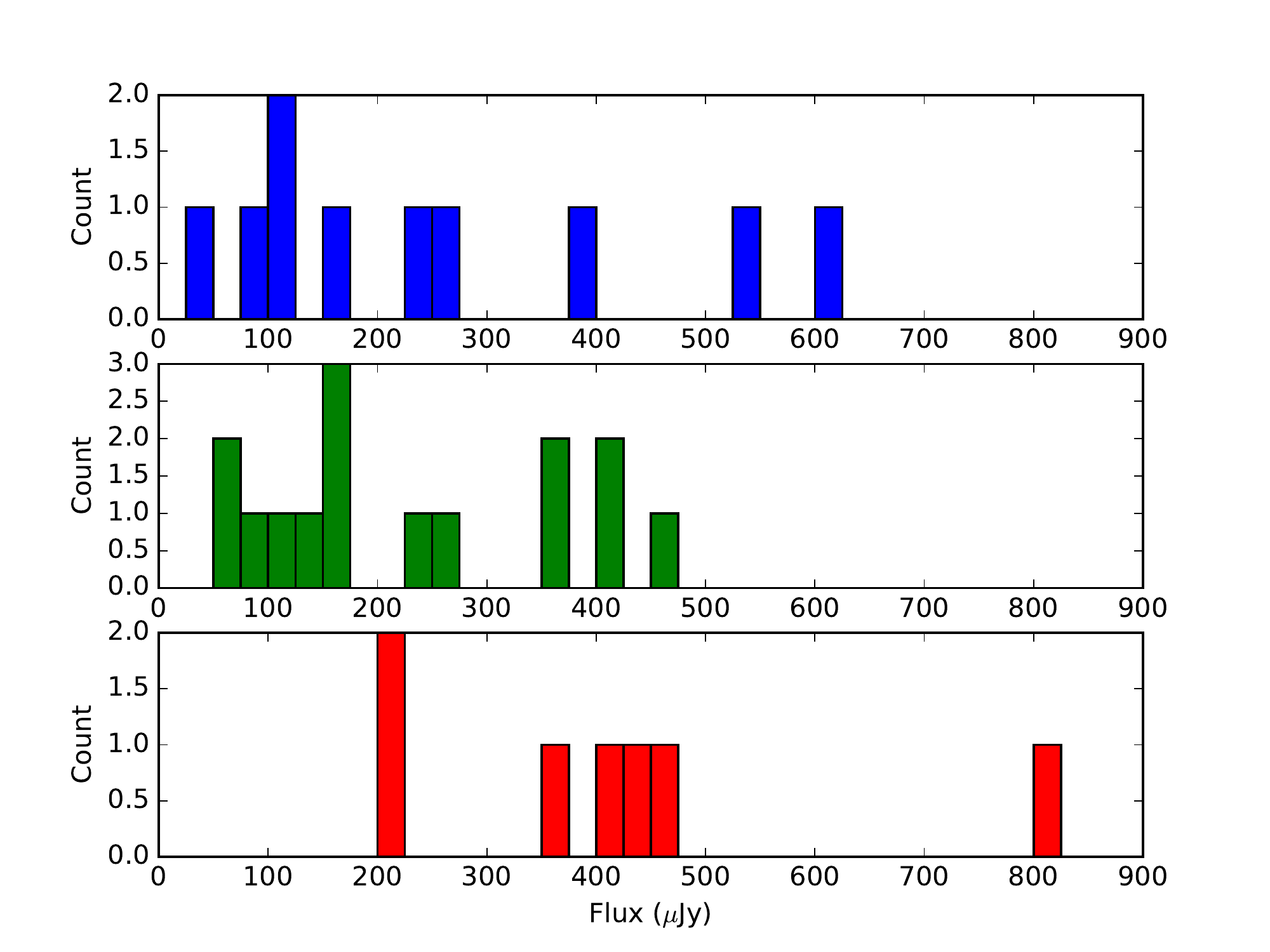}{3.3in}{Figure 1. Flux histograms.  From top
    to bottom, the panels are the C-band, X-band, and K-band flux
    densities, respectively.  The flux densities of AE~Aqr, being $>5$
    mJy, are excluded.}
\end{figure}
AE~Aqr and AR~UMa show radio emission at all frequencies (C-, X-, \&
K-bands) and at both epochs.  Whereas AM~Her shows emission at all
frequencies at only a single epoch.  The remaining sixteen sources
show emission at no more than two frequencies.  All CVs show variable
emission at some level, even the three strong sources; AE~Aqr, AM~Her,
and AR~UMa.  The variability of the radio flux at different epochs and
frequencies is characteristic of short timescale flares.  For example,
the polar AM~Her shows faint (114 $\mu$Jy) C-band emission during the
first observation on 2013-12-02 and moderate emission (154, 243, \&
818 $\mu$Jy) at three frequencies (C, X, \& K, respectively) during
the second observation 34 days later on 2014-01-06.





The small sample (19) of detections limits the comparison of the
frequency dependence of the radio emission to two subclasses, polars
and non-polars.  For the polars, the number of detections at the C-,
X-, and K-bands are 7, 12, and 7, respectively.  For the non-polars
with the exclusion of AE~Aqr, the numbers are respectively 3, 3, and
0.  This suggests that the polars are more likely to emit at higher
frequencies ($> 22$ GHz) compared to the non-polars.  We note that,
except for AE~Aqr, these observations are the first detections of
polars at 22 GHz.



A notable feature of Table 3 is the fourteen detections that show
essentially 100\% circularly polarized emission (i.e., a source is
only seen in the left-hand or right-hand circular polarization image),
whereas the remaining five detections show no (linear and circularly)
polarized emission.  In addition the timescale of the emission is
short, $<5$ minutes.  These two properties are characteristic of
electron-cyclotron maser emission like that of the ten minute flare of
AM~Her.  Four of these sources show polarized emission at 22 GHz.  The
other five radio sources show no (or weakly) polarized emission.
Three of them are the well-known MCVs, AM~Her, AR~UMa, and AE~Aqr.
These results suggest that MCVs might be divided into two classes of
radio emitters: emission dominated by weakly polarized gyrosynchrotron
radiation and by highly polarized electron-cyclotron maser emission.

\section{Conclusions}

The current survey indicates that MCVs are highly variable radio
sources that can emit at frequencies as high as 22 GHz via
gyrosynchrotron and electron-cyclotron maser emission.  The data
suggest that the polars are more likely to emit at higher frequencies
than the non-polars, and MCVs can be divided into two classes of radio
emitters: those dominated by weakly polarized gyrosynchrotron emission
and those by highly polarized electron-cyclotron maser emission.

Paper two of this survey presents a more in-depth analysis of these
data by investigating the two radiation mechanisms and the possibility
of two classes of radio emitting MCVs.  Paper three of this survey
will present high-resolution VLBI observations of MCVs, including the
sources detected in this paper.

\acknowledgments

P.E.B. and P.A.M. acknowledge the guidance and inspiration of their
late mentor and colleague G. Chanmugam, who began the search for radio
emission from magnetic cataclysmic variables over 35 years ago.  The
National Radio Astronomy Observatory is a facility of the National
Science Foundation operated under cooperative agreement by Associated
Universities, Inc.  This research has made use of the SIMBAD database,
operated at CDS, Strasbourg, France \citep{weng00}.

\end{document}